\documentclass[aps,prb,a4paper,superscriptaddress,twocolumn,showpacs,amsmath,amssymb]{revtex4} 
\usepackage{graphicx,epsfig} 
\usepackage{dsfont,bbm}
\allowdisplaybreaks
\begin{document} 

\newcommand{\s}{\sigma}
\newcommand{\up}{\uparrow}
\newcommand{\down}{\downarrow}
\newcommand{\h}{\mathcal{H}}
\newcommand{\g}{\mathcal{G}^{-1}_0}
\newcommand{\D}{\mathcal{D}}
\newcommand{\A}{\mathcal{A}}
\newcommand{\K}{\textbf{k}}
\newcommand{\Q}{\textbf{q}}
\newcommand{\T}{\tau_{\ast}}
\newcommand{\io}{i\omega_n}
\newcommand{\eps}{\varepsilon}
\newcommand{\+}{\dag}
\newcommand{\su}{\uparrow}
\newcommand{\giu}{\downarrow}
\newcommand{\0}[1]{\textbf{#1}}
\newcommand{\ca}{c^{\phantom{\dagger}}}
\newcommand{\cc}{c^\dagger}
\newcommand{\da}{d^{\phantom{\dagger}}}
\newcommand{\dc}{d^\dagger}
\newcommand{\be}{\begin{equation}}
\newcommand{\ee}{\end{equation}}
\newcommand{\bea}{\begin{eqnarray}}
\newcommand{\eea}{\end{eqnarray}}
\newcommand{\ba}{\begin{eqnarray*}}
\newcommand{\ea}{\end{eqnarray*}}
\newcommand{\dagga}{{\phantom{\dagger}}}
\newcommand{\bR}{\mathbf{R}}
\newcommand{\bQ}{\mathbf{Q}}
\newcommand{\bq}{\mathbf{q}}
\newcommand{\bqp}{\mathbf{q'}}
\newcommand{\bk}{\mathbf{k}}
\newcommand{\bh}{\mathbf{h}}
\newcommand{\bkp}{\mathbf{k'}}
\newcommand{\bp}{\mathbf{p}}
\newcommand{\bRp}{\mathbf{R'}}
\newcommand{\bx}{\mathbf{x}}
\newcommand{\by}{\mathbf{y}}
\newcommand{\bz}{\mathbf{z}}
\newcommand{\br}{\mathbf{r}}
\newcommand{\Ima}{{\Im m}}
\newcommand{\Rea}{{\Re e}}
\newcommand{\Pj}[2]{|#1\rangle\langle #2|}
\newcommand{\ket}[1]{\vert#1\rangle}
\newcommand{\bra}[1]{\langle#1\vert}
\newcommand{\fract}[2]{\frac{\displaystyle #1}{\displaystyle #2}}
\newcommand{\Av}[2]{\langle #1|\,#2\,|#1\rangle}
\newcommand{\Avs}[1]{\langle \,#1\,\rangle_0}
\newcommand{\eqn}[1]{(\ref{#1})}
\newcommand{\Tr}{\mathrm{Tr}}
\title{Superconductivity in the doped bilayer Hubbard model.}
\author{Nicola Lanat\`a}
\affiliation{International School for Advanced Studies (SISSA), and CRS Democritos, CNR-INFM,
Via Beirut 2-4, I-34014 Trieste, Italy} 
\author{Paolo Barone}
\affiliation{International School for Advanced Studies (SISSA), and CRS Democritos, CNR-INFM,
Via Beirut 2-4, I-34014 Trieste, Italy} 
\author{Michele Fabrizio} 
\affiliation{International School for Advanced Studies (SISSA), and CRS Democritos, CNR-INFM,
Via Beirut 2-4, I-34014 Trieste, Italy}
\affiliation{The Abdus Salam International Centre for Theoretical Physics 
(ICTP), P.O.Box 586, I-34014 Trieste, Italy} 
\date{\today} 
\begin{abstract}
We study by the Gutzwiller approximation the melting of the valence bond crystal phase of a bilayer Hubbard model 
at sufficiently large inter-layer hopping. We find that a superconducting domain, with order parameter $d_{z^2-r^2}$, $z$ being 
the inter-layer direction and $r$ the intra-layer one, is stabilized variationally close to the 
half-filled non-magnetic Mott insulator. Superconductivity exists at half-filling just at the border of the Mott transition 
and extends away from half-filling into a whole region till a critical doping, beyond which it gives way to 
a normal metal phase. This result suggests that superconductivity should be unavoidably met by liquefying a valence bond crystal, 
at least when each layer is an infinite coordination lattice and the Gutzwiller approximation becomes exact.
Remarkably, this same behavior is well established in the other extreme of two-leg Hubbard ladders, showing it might be of 
quite general validity. 
\end{abstract}

\maketitle

\section{Introduction}

Since its original formulation in the early 60th's, the Gutzwiller variational approach~\cite{Gutzwiller1,Gutzwiller2,Gutzwiller3} 
has proved to be one of the simplest yet effective tools to deal with correlated electron systems. 

The basic idea of the method is to modify variationally the weights of local electronic configurations with respect 
to an uncorrelated wavefunction $\vert\Psi_0\rangle$, for which Wick's theorem holds, 
according to the local interaction terms. This is accomplished by means of the variational wavefunction:
\be
|\Psi_G\rangle = \mathcal{P}\vert\Psi_0\rangle=\prod_\bR\,\mathcal{P}_\bR\,\vert\Psi_0\rangle,\label{GWF}
\ee
where $\mathcal{P}_\bR$ is an operator acting on the local Hilbert space of the unit cell $\bR$. Both the 
uncorrelated wavefunction $\vert\Psi_0\rangle$ and the operators $\mathcal{P}_\bR$ must be determined variationally by minimizing 
the average energy. In general the average energy can be calculated only numerically~\cite{Sorella-VMC} but in the limit of 
infinite coordination lattices, where a lot of simplifications intervene~\cite{Muller} that allow for an explicit analytical 
expression~\cite{Metzner-Vollhardt-PRL,Metzner-Vollhardt-PRB,Gebhard}. 
This is rigorously valid only in infinite coordination lattices, nevertheless it is commonly used also in finite 
coordination ones, what is refereed to as the Gutzwiller approximation because in a single band model it happens to coincide 
with the approximation introduced by Gutzwiller himself~\cite{Gutzwiller3}.   

In spite of its simplicity, many important concepts in strongly correlated electron systems have originated 
from Gutzwiller variational calculations or, which is equivalent~\cite{Kotliar-Ruckenstein,quaquaraqua}, 
from slave-boson mean-field theory~\cite{Kotliar-Ruckenstein,Georges}. We just mention the famous Brinkmann-Rice 
scenario~\cite{Brinkman&Rice} of the Mott transition. Therefore, even though more rigorous approaches have been developed 
meanwhile, like DMFT~\cite{DMFT} or LDA+U~\cite{LDA+U}, there has been a continuous effort towards improving 
the original Gutzwiller wavefunction in finite dimensions~\cite{Capello1}, and extending the Gutzwiller approximation 
to account for the exchange interaction in multi-orbital models~\cite{Gebhard,Bunemann,Attaccalite,mio}, for 
the electron-phonon coupling~\cite{Barone1}, for interfaces effects~\cite{Borghi}, 
and also for more {\sl ab-initio} ingredients~\cite{Fang}. A reason for this perseverance is that 
the Gutzwiller wavefunction and approximation are so simple and flexible to be adapted to many different situations and provide 
without big numerical efforts reasonable results. 

In its simplest formulation, Eq.~\eqn{GWF}, the form of intersite correlations within the Gutzwiller wavefunction are  
controlled solely by the uncorrelated wavefunction $\vert\Psi_0\rangle$. This aspect should not be problematic if the main interest 
is in the gross features near a Mott transition or when a Hartree-Fock Slater determinant gives already a 
reasonable description of the actual ground state, which can only be improved by applying the operator 
$\mathcal{P}$. However, there are interesting cases where new types of correlations may arise near a Mott transition that 
are not explicitly present in the Hamiltonian. A known example are the $d$-wave superconducting fluctuations that are believed 
to emerge in the single band Hubbard model on a square lattice 
close to the half-filled antiferromagnetic Mott insulator~\cite{Scalapino}, and 
which are often invoked to explain high T$_c$ superconductivity. A simple way to justify the emergence of superconducting 
fluctuations is to take the large $U$ limit of the Hubbard model, which is known to correspond in the low energy sector 
to the $t$-$J$ model.  Here, the antiferromagnetic exchange $J$ provides an explicit attraction 
in the inter-site singlet channel. This reflects the tendency of neighboring sites to form spin singlets, which 
turns into a true antiferromagnetic long-range order at half-filling but may mediate superconductivity upon doping. 
Indeed, the Gutzwiller approximation and equivalently the slave-boson mean-field theory applied to the $t$-$J$ model  
do stabilize a $d$-wave superconducting phase away from half-filling~\cite{Kotliar&Liu}, a result supported by direct 
numerical optimizations of $|\Psi_G\rangle$, with $\mathcal{P}_\bR$ projecting out doubly occupied sites and  
$\vert\Psi_0\rangle$ a $d$-wave BCS-like wavefunction~\cite{Gros-PRB,Paramekanti,Sandro&Dagotto,Vanilla}. 
However, the simplest Gutzwiller approximation in the pure Hubbard model away from half-filling 
does not stabilize any superconducting phase, just because the on-site repulsion $U$ does not couple directly to the 
$d$-wave superconducting parameter. A way to improve the wavefunction allowing for inter-site spin-singlet correlations 
could be using an enlarged non-primitive unit cell, with $\mathcal{P}_\bR$ in \eqn{GWF} acting on a cluster of sites. 
With this choice the wavefunction $|\Psi_G\rangle$ breaks explicitly lattice translational symmetry, so that one 
should properly modify the variational scheme not to get spurious results, just like any 
other cluster technique~\cite{CDMFT-Lichtenstein,CDMFT-Potthoff,CDMFT-Senechal,CDMFT-Kotliar,CDMFT-Jarrell,Arrigoni}. 

Alternatively, one might consider different models that are manageable with the simple wavefunction \eqn{GWF} and 
which are expected to have a physical behavior similar to the one looked for in the Hubbard model on a square lattice.  
One case by now well known is that of two coupled 
Hubbard or $t$-$J$ chains. At half-filling, both models are non-magnetic Mott insulators~\cite{Strong&Millis,shura,Balents&Fisher}. 
The insulating phase is a kind of short-range resonating valence bond (RVB) spin-liquid~\cite{Scalapino-RVB}, i.e. 
a spin-gaped state without any symmetry breaking. Actually this state is adiabatically connected 
to the trivial insulator for very large inter-chain coupling, which is a collection of inter-chain dimers, what can be 
denoted as a valence bond (VB) crystal. Away from half-filling, dominant 
superconducting fluctuations arise~\cite{2chain,Balents&Fisher,Schulz}, with a two-chain analog of a two-dimensional $d$-wave symmetry. 
The emergence of strong superconducting fluctuations appears here as the natural fate of doping the half-filled 
VB Mott insulator~\cite{Scalapino-RVB}, realizing in one-dimension the RVB superconductivity scenario proposed 
by Anderson~\cite{PWA} in the early days after the discovery of high T$_c$ superconductivity. An immediate question that arises is 
whether the above one-dimensional behavior survives in higher-dimensions, namely how robust is the two-chain 
RVB scenario upon increasing dimensionality. This is actually the content of the present work. 

As a matter of fact, this question has been already addressed several times in connection with 
high T$_c$ superconductors, specifically analyzing a bilayer Hubbard model by various techniques, including 
quantum Monte Carlo~\cite{J-Jperp-1,Scalettar,dos-Santos, J-Jperp-2,Referee} (QMC) and DMFT~\cite{Monien,Okamoto}. 
In section~\ref{The Model} we shall discuss more in detail these early works while introducing the model. 
More recently, the same problem has been studied 
at half-filling by an improved Gutzwiller approximation~\cite{mio}, which we present in section~\ref{Method} together with 
a further improvement that we use here to extend that analysis away from half-filling. The results are presented in 
section~\ref{Results}, while section~\ref{Conclusions} is devoted to concluding remarks.

\section{The model}\label{The Model}

Throughout this work we shall be interested in a bilayer Hubbard model described by the following Hamiltonian
\bea
\mathcal{H}&=&-\sum_{\bR\bRp}\,\sum_{i=1}^2\,\sum_\sigma\,
t_{\bR\bRp}\,\cc_{\bR,i\sigma}\ca_{\bRp,i\sigma}+ H.c.\nonumber\\
&&+\frac{U}{2}\sum_\bR\,\sum_{i=1}^2\,(n_{\bR,i}-1)^2 \nonumber\\
&&-t_\perp\,\sum_\bR\,\sum_\sigma\,\left(\cc_{\bR,1\sigma}\ca_{\bR,2\sigma} + H.c.\right)\nonumber \\
&=& \sum_{\bk\sigma}\,\sum_{i=1}^2\,\epsilon(\bk)\,c^\dagger_{\bk,i\sigma}c^\dagga_{\bk,i\sigma}\nonumber\\
&&+\frac{U}{2}\sum_\bR\,\sum_{i=1}^2\,(n_{\bR,i}-1)^2 \nonumber\\
&&-t_\perp\,\sum_{\bk\sigma}\,\left(\cc_{\bk,1\sigma}\ca_{\bk,2\sigma} + H.c.\right)\nonumber \\
&&\equiv \mathcal{H}_{hop} + \mathcal{H}_U + \mathcal{H}_\perp, \label{ham}
\eea
where $t_\perp>0$, $c^\dagger_{\bR,i\sigma}$ and $c^\dagga_{\bR,i\sigma}$ create and annihilate, respectively, an electron 
at site $\bR$ in plane $i=1,2$ with spin $\sigma$, 
$n_{\bR,i}=\sum_\sigma\,\cc_{\bR,i\sigma}\ca_{\bR,i\sigma}$ is the local occupation on layer $i$,  
and $U$ is the Hubbard repulsion on each lattice site. 
In order to study the doped system it is more convenient to work in the grand-canonical
ensemble adding a chemical potential term $-\mu\sum_{\bR,i} n_{\bR,i}$ to the model Hamiltonian (\ref{ham}).
The particle number is then controlled by tuning $\mu$. In Eq.~\eqn{ham} 
$c^\dagger_{\bk,i\sigma}$ creates an electron in layer $i$ and spin $\sigma$ 
with momentum $\bk$, and $\epsilon(\bk)\in [-D,D]$ is the intra-layer dispersion 
in momentum space, where $D$ is half the bandwidth that will be our unit of energy. 
The non-interacting part of the Hamiltonian is better rewritten introducing the bonding ($e$) 
and antibonding ($o$) combinations
\ba
c^\dagger_{\bk,e\sigma} &=& \fract{1}{\sqrt{2}}\left(c^\dagger_{\bk,1\sigma} + c^\dagger_{\bk,2\sigma}\right),\\
c^\dagger_{\bk,o\sigma} &=& \fract{1}{\sqrt{2}}\left(c^\dagger_{\bk,1\sigma} - c^\dagger_{\bk,2\sigma}\right),
\ea
through which
\be
\mathcal{H}_{hop} + \mathcal{H}_\perp = 
\sum_{\bk\sigma}\,\sum_{a=e,o}\,\epsilon_a(\bk)\,c^\dagger_{\bk,a\sigma}c^\dagga_{\bk,a\sigma},\label{Ham-e-o}
\ee
where $\epsilon_e(\bk) = \epsilon(\bk) - t_\perp\in \left[-D-t_\perp,D-t_\perp\right]$ 
and $\epsilon_o(\bk) = \epsilon(\bk) + t_\perp\in \left[-D+t_\perp,D+t_\perp\right]$ are, respectively, the bonding and 
antibonding band dispersions. 

If $U=0$ and the density is one electron per site, half-filling, the model describes a metal until the two bands overlap, 
i.e. $t_\perp \leq D$, and a band insulator otherwise. 

For $U\gg D+t_\perp$, the model becomes equivalent 
to two Heisenberg planes coupled to each other by an inter-plane antiferromagnetic exchange $J_\perp=4t_\perp^2/U$. 
If each plane is a square lattice with only  
nearest neighbor hopping $t$, hence $D=4t$, each Heisenberg model is characterized by a nearest neighbor antiferromagnetic 
exchange $J=4t^2/U$. This model has been studied in detail by quantum Monte Carlo~\cite{J-Jperp-1,J-Jperp-2} 
and it is known to have a quantum critical point that separates a 
a Ne\`el antiferromagnet, for $J_\perp\leq 2.5520\,J$, from a gaped spin-liquid phase, for larger $J_\perp$. 
The latter can be interpreted as a kind of VB crystal, each bond being an inter-layer singlet, adiabatically connected to the 
band insulator at $U=0$. 
In terms of the hopping parameters of the original Hubbard bilayer, the critical point should correspond to 
$(t_\perp/t)_c = \sqrt{2.5220} \simeq 1.5881$. This value is in good agreement with direct QMC  
simulations of the Hubbard bilayer~\cite{Scalettar,Referee}, which find $(t_\perp/t)_c\simeq 1.5$ to 2. 
According to these results, when $1.6\leq (t_\perp/t)\leq 4$ one could start at $U=0$ with a metallic phase, and, upon increasing $U$, 
find a direct transition into the VB Mott insulator.
However, the story must become more complicated if the $U=0$ Fermi surface at half-filling has nesting at the edge of the 
Brillouin zone, as it happens for a square lattice with only nearest neighbor hopping. 
In this case, the $U=0$ and $t_\perp < 4t=D$ metal has a Stoner 
instability towards Ne\`el antiferromagnetism for arbitrary small $U$, so that it is {\sl a priori} not obvious that one could 
find any direct metal to VB Mott insulator transition. In reality, both cluster DMFT~\cite{Okamoto} and QMC  
simulations find evidence that such a transition does exist. Nevertheless, one may always bypass this problem assuming that the 
intra-layer hopping is such as not to lead to any nesting, the latter being more an accident than the rule in realistic 
systems. In this case, which we will implicitly assume hereafter, it is safe to believe that a direct transition at half-filling 
from a metal to a VB Mott insulator does exist. 

Within this scenario, the melting of the VB crystal into a metallic phase can therefore occur either by doping away from half-filling
but also upon decreasing $U$ below the Mott transition, still keeping half-filled density. In the latter case, a recent 
study~\cite{mio} has shown that, within the Gutzwiller approximation, the VB crystal first turns into a superconducting phase 
that eventually gives way to a normal metal upon further decreasing $U$. This finding supports the RVB superconductivity 
scenario~\cite{PWA} and shows that the one-dimensional behavior persists in higher dimensions. It also agrees with the indication of an 
enhanced pairing susceptibility obtained in earlier studies by QMC~\cite{Bulut,dos-Santos}. However the lowest temperatures 
attainable so far by QMC are still above the eventual superconducting critical temperature, so that the existence 
of a true superconducting phase at half-filling is numerically still an open issue. 
DMFT calculations, that could in principle be carried out at zero temperature, 
was performed~\cite{Monien,Okamoto} but did not search explicitly for any superconducting phase. 

Away from half-filling, QMC indications of enhanced pairing fluctuations are more convincing~\cite{dos-Santos,Referee}, 
although the existence of a superconducting phase at low temperature is still uncertain~\cite{dos-Santos}. This 
makes it worth addressing this issue by the Gutzwiller approximation, which is not as rigorous as QMC but at least can 
provide results at zero temperature.

\section{The method}\label{Method}

In order to study the bilayer Hubbard model \eqn{ham} away from half-filling we adopt the Gutzwiller approximation scheme developed 
in Refs.~\onlinecite{mio} to deal with the same model at half-filling. The variational wavefunction that we use has the 
form as in Eq.~\eqn{GWF} where 
\begin{itemize}
\item[1.] $\mathcal{P}_\bR$ acts on the full Hilbert space that includes site $\bR$ in layer 1 and site $\bR$ in layer 2;
\item[2.] $|\Psi_0\rangle$ is allowed to be a BCS-wavefunction with singlet order parameter in the channel 
$c^\dagger_{\bR,1\up}c^\dagger_{\bR,2\down} + c^\dagger_{\bR,2\up}c^\dagger_{\bR,1\down}$.
\end{itemize}  
The most general expression for $\mathcal{P}_\bR$ is:
\be
\mathcal{P}_\bR = \sum_{\Gamma_1,\Gamma_2}\,\lambda(\bR)_{\Gamma_1 \Gamma_2}\,\Pj{\Gamma_1,\bR}{\Gamma_2,\bR},\label{PR}
\ee
where each state $\ket{\Gamma_i,\bR}$ denotes a local two-site electronic configuration,
and the matrix $\lambda(\bR)$ has to be variationally determined. 
Average values of operators on the wavefunction (\ref{GWF}) can be analytically computed in infinite-coordination
lattices provided the following constraints are satisfied by $\mathcal{P}_\bR$~\cite{Gebhard,Attaccalite,mio}:
\bea
\Av{\Psi_0}{\mathcal{P}^\dagger_\bR\,\mathcal{P}^\dagga_\bR} &=& 1,\label{uno}\\
\Av{\Psi_0}{\mathcal{P}^\dagger_\bR\,\mathcal{P}^\dagga_\bR\,\mathcal{C}_{\bR}} &=& 
\Av{\Psi_0}{\mathcal{C}_{\bR}},\label{due}
\eea 
where $\mathcal{C}_{\bR}$ is the local single-particle density-matrix operator with elements $\cc_{\bR,\alpha}\ca_{\bR,\beta}$ and
$\cc_{\bR,\alpha}\cc_{\bR,\beta}$, $\alpha$ and $\beta$ labeling single-particle states (both layer and spin indices) 
and $\cc_{\bR,\alpha}$($\ca_{\bR,\alpha}$) creating(annihilating) 
an electron in state $\alpha$ at site $\bR$. Expectation values of local operators are then
computed straightforwardly~\cite{mio} as  
$\Avs{\mathcal{P}^\dagger \mathcal{O}_{\bR} \mathcal{P}}=\Avs{\mathcal{P}^\dagger_{\bR}\mathcal{O}_{\bR}\mathcal{P}_{\bR}}$ 
(hereafter $\Avs{...}$ denotes averages on the uncorrelated wavefunction $\vert\Psi_0\rangle$, which can be easily computed by means 
of Wick's theorem). When calculating the average of the inter-site density matrix, one finds that the physical single-fermion 
operator acting on $|\Psi_G\rangle$ is effectively replaced by a renormalized one acting on $|\Psi_0\rangle$ according to: 
\be
 c^\dagger_{\bR,\alpha} \rightarrow \sum_{\beta } R(\bR)_{\alpha\beta}\,c^\dagger_{\bR,\beta} 
+ \sum_{\beta }  Q(\bR)_{\alpha\beta}\,\ca_{\bR,\beta},\label{Z-qp}
\ee
where the renormalization matrices $R$ and $Q$ are determined by inverting the following set of equations\cite{mio}:
\bea
&&\Avs{\mathcal{P}^\dagger_{\bR}\cc_{\bR,\alpha}\mathcal{P}_{\bR}\ca_{\bR,\beta}}
=\sum_\gamma\,R(\bR)_{\alpha\gamma}\, \Avs{\cc_{\bR,\gamma}
\ca_{\bR,\beta}} \nonumber\\
&&~~~~~~~~~ + \sum_\gamma\, Q(\bR)_{\alpha\gamma}\, \Avs{\ca_{\bR,\gamma}
\ca_{\bR,\beta}} ,\\
&& \Avs{\mathcal{P}^\dagger_{\bR}\cc_{\bR,\alpha}\mathcal{P}_{\bR}\cc_{\bR,\beta}}
=\sum_\gamma\,R(\bR)_{\alpha\gamma}\,\Avs{\cc_{\bR,\gamma}
\cc_{\bR,\beta}} \nonumber\\
&&~~~~~~~~~+ \sum_\gamma\, Q(\bR)_{\alpha\gamma}\, \Avs{\ca_{\bR,\gamma}
\cc_{\bR,\beta}} .
\eea
In Ref.~\onlinecite{mio}, a different notation was used for the matrices $R$ and $Q$, namely $R=\sqrt{Z}$ and 
$Q=\sqrt{\Delta}$. In order not to generate any confusion with the definition of a square root of a matrix, and 
also for keeping more explicit the connection with slave-boson mean field theory~\cite{Georges}, we have preferred 
here to use $R$ and $Q$.  
Despite the considerable simplification introduced by the infinite-coordination
limit, the variational problem remains still a difficult task to deal with because of the large size  
of the local Hilbert space, which contains 16 states so that $\lambda$ spans in principle $16\times 16$ matrices.  

A further simplification can be achieved with a proper choice of the basis set spanning the local Hilbert space. 
This can be done, for instance, by using from the
beginning the natural basis, i.e., the single-particle basis which diagonalizes the variational density matrix
$\Avs{\mathcal{C}_{\bR}}$\cite{mio}. An alternative and more efficient approach consists\cite{nostro} in defining the local operator
$\mathcal{P}_\bR$ in a mixed-basis representation, namely
expressing $\ket{\Gamma_1,\bR}=\ket{\Gamma,\bR}$ in Eq.~(\ref{PR}) in the
original basis defined by the model Hamiltonian and assuming
that $\bra{\Gamma_2,\bR}=\bra{\{\bar{n}_\alpha\},\bR}$ are Fock states in the natural basis, identified by the occupation numbers 
$\bar{n}_\alpha = 0,1$. With this choice, one can use as variational parameters just the eigenvalues of the density matrix, because the
unitary transformation that relates the natural-basis operators $\da_{\bR,\alpha}$ to the original ones $\ca_{\bR,\alpha}$ 
needs not to be known explicitly. This simplifies considerably all calculations.  
In the mixed original-natural basis representation one introduces a new matrix
\be
\phi(\bR) = \lambda(\bR)\,\sqrt{P^0(\bR)},\label{phi}
\ee
where $\lambda(\bR)$ is the variational matrix in the mixed-basis representation and $P^0(\bR)$  is the
uncorrelated occupation-probability matrix, with elements 
\ba
P^0_{\{\bar{n}_\alpha\},\{\bar{m}_\alpha\}}(\bR) \equiv&& \Avs{\vert \{\bar{m}_\alpha\},\bR\rangle\langle
\{\bar{n}_\alpha\},\bR\vert} \\
&& = \delta_{\{\bar{n}_\alpha\}\{\bar{m}_\alpha\}}\,P^0_{\{\bar{n}_\alpha\}}(\bR),
\ea
being  
\be
P^0(\bR)_{\{\bar{n}_\alpha\}} =\prod_{\alpha}\, \left(n^0(\bR)_{\alpha}\right)^{\bar{n}_{\alpha}}\,
\left(1-n^0(\bR)_{\alpha}\right)^{1-\bar{n}_{\alpha}},\label{Pn0}
\ee
and $n^0(\bR)_{\alpha}$ the eigenvalues of the density matrix (to be variationally determined). In terms of $\phi$,  
the constraints to be imposed on the Gutzwiller wavefunction can be recast as\cite{nostro}:
\bea
\Tr\left(\phi^\dagger\phi\right) &=& 1, \label{uno_n}\\
\Tr\left(\phi^\dagger\phi \, \dc_\alpha \da_\beta  \right) &=& \delta_{\alpha,\beta} \, n^0_\alpha,\label{due_n}\\
\Tr\left(\phi^\dagger\phi \, \dc_\alpha \dc_\beta  \right) &=& 0,\label{tre_n}
\eea
where, to simplify notations, we dropped the site-label
$\bR$, and we also introduced matrix representations of the fermionic single-particle operators. The average of local operators 
and the renormalization factors acquire very simple expressions: 
\bea
\Avs{\mathcal{P}^\dagger
\mathcal{O} \mathcal{P}} &=& \Tr\,\bigg(\phi^\dagger\mathcal{O}\phi\bigg),\label{locav}\\
R_{\alpha\beta}
&=&  \frac{1}{\sqrt{n^0_{\beta}(1-n^0_{\beta})}}\;
\Tr\,\bigg(\phi^{\dagger}\,\cc_{\alpha}\,\phi\, \da_{\beta}\bigg), \label{Z-new}\\
Q_{\alpha\beta}
&=&  \frac{1}{\sqrt{n^0_{\beta}(1-n^0_{\beta})}}\;
\Tr\,\bigg(\phi^{\dagger}\,\cc_{\alpha}\,\phi\, \dc_{\beta}\bigg).
\label{D-new}
\eea
Through Eqs.~(\ref{Z-new}) and (\ref{D-new}) 
single-particle fermionic operators are automatically mapped into renormalized operators in the natural basis,
and Eq.~(\ref{Z-qp}) is replaced by:
\be
\cc_{\alpha} \rightarrow \sum_{\beta } R_{\alpha\beta}\,\dc_{\beta} + \sum_{\beta } Q_{\alpha\beta}\,\da_{\beta}.\label{Z-qpnew}
\ee
Note the presence of the latter term in the rhs of Eq.~\eqn{Z-qpnew}, which makes it possible that a creation 
operator in the original representation turns into an annihilation operator in the natural one. Its existence is a direct 
consequence of allowing $|\Psi_0\rangle$ to span also BCS-like wavefunctions and/or $\mathcal{P}_\bR$ 
to couple states with different particle numbers. Should $|\Psi_0\rangle$ describe a normal metal and $\mathcal{P}_\bR$ be diagonal 
in the particle number, $Q_{\alpha\beta}$ would be strictly zero, as was the case in Ref.~\onlinecite{nostro}. Therefore 
Eqs.~\eqn{Z-new}, \eqn{D-new} and \eqn{Z-qpnew} extend Eqs.~(A10) and (A18) of Ref.~\onlinecite{nostro} to the more general case
in which superconductivity is allowed.   

Practically, it is convenient~\cite{nostro} to generate variational matrices $\phi$ that directly 
satisfy Eqs.~\eqn{uno_n}-\eqn{tre_n} hence univocally determine the parameters $n^0_\alpha$, 
and only after impose, by proper Lagrange multipliers, that the uncorrelated $|\Psi_0\rangle$ 
has an average local density matrix with eigenvalues $n^0_\alpha$.
We end mentioning that the elements $ \phi_{\Gamma,\{\bar{n}\}} = 
\lambda_{\Gamma \{\bar{n}\}}\,\sqrt{P^0_{\{\bar{n}\}}}$ of the matrix (\ref{phi})
correspond to the slave-boson saddle-point values within the mean-field scheme recently introduced 
by Lechermann and coworkers~\cite{Georges}.

It may happen that, in spite of all the above simplifications, the variational space thus
generated  is still unnecessarily large. For instance, if one looks for a
variational wavefunction that preserves particle number, all the elements of $\lambda$ connecting subspaces of the 
local Hilbert space with different particle numbers should be identically zero. Therefore it would be desirable to specialize 
the general procedure sketched above in such a
way that symmetries can be built in the variational wavefunction from the onset.  
In general, given a symmetry group $G_0$ that one would like to enforce, 
we must require, in addition to \eqn{uno} and \eqn{due}, that 
\be
[\mathcal{P}_\bR,G_0]=0,
\label{sym}
\ee
However, in the mixed representation there may be some symmetry operations that can not be defined without 
an explicit knowledge of the natural basis in terms of the original one, which would make the whole method much 
less efficient. If one decides not to implement these symmetries, but only those, symmetry group $G\subset G_0$, 
that commute with the most general unitary transformation $U$ connecting original and 
natural basis, i.e.  
\[
\left[U,G\right]=0,
\]
compatibly with the variational ansatz, the 
above described variational method can be still used with the following modification.  

Let us assume this case and define a unitary operator $V$ that transforms the Fock states in the original basis 
into states that decompose the local Hilbert space in irreducible representations of the group $G$. We define $\bar{G}$ the 
representation of $G$ in such a basis. Because of our choice of the subgroup $G$,  
$V$ does right the same job even in the natural basis, although this is unknown. 
Since the trace is invariant under unitary transformations, 
all formulas from \eqn{uno_n} to \eqn{D-new} remain the same even if the variational matrix $\phi$ and the matrix representation 
of the single fermion operators are defined in the states of the irreducible representation, both in the 
original and natural basis, with the additional symmetry constraint 
\be
\left[\phi,\bar{G}\right] = 0, \label{sym-2}
\ee
which follows from \eqn{sym}. We note that the matrix representation of a single-fermion operator in these states is 
readily obtained once $V$ is known, and is trivially the same for both original and natural operators. 
Therefore it is sufficient to create and store it at the beginning of any calculation. 

As an example, which is directly pertinent to this work, let us consider $G$ the group of spin $SU(2)$ transformations. 
In this case an irreducible representation is readily obtained and consists of states with fixed total spin $S$ and 
its $z$-projection $S_z$, of the general form $|\Gamma,S,S_z\rangle$, where $\Gamma$ serves as an additional label to 
distinguish between different states with same $S$ and $S_z$ in the original representation. The operator $V$ is thus 
the unitary transformation that connects Fock states in the original basis, $|\{n_\alpha\}\rangle$ to the 
states $|\Gamma,S,S_z\rangle$,
\[
V\,:\; |\{n_\alpha\}\rangle \rightarrow |\Gamma,S,S_z\rangle.
\]
We use the same $V$ to generate from the Fock states in the natural basis, which we remind is and remains unknown, 
the states $|\bar{\Gamma},S,S_z\rangle$. It follows that, in order to 
preserve full spin $SU(2)$ symmetry, 
\be
\phi_{\Gamma,S,S_z;\bar{\Gamma},\bar{S},\bar{S}_z} = \delta_{S\bar{S}}\;
\delta_{S_z \bar{S}_z}\;\phi_{\Gamma,S;\bar{\Gamma},S},\label{phi-to-preserve}
\ee
is a block matrix. 

\section{Results}\label{Results}

Let us turn now to study the bilayer Hubbard model \eqn{ham} away from half-filling. 
As mentioned, the filling is controlled by a chemical potential term $-\mu\sum_{\bR\, i} n_{\bR,i}$ that we add to the 
Hamiltonian \eqn{ham}. We search for a variational solution 
that allows for singlet superconductivity and doesn't break spin-$SU(2)$ symmetry. In this case, the unitary transformation that 
would connect original to natural bases, would leave spin $SU(2)$ generators invariant, so that 
the method described in the previous section is applicable.  

We already said that the Gutzwiller operator 
$\mathcal{P}_\bR$ in \eqn{PR} acts on the whole local Hilbert space of two sites, $\bR$ in layer 1 and $\bR$ in layer 2.
The variational energy to be 
minimized is then the sum of two terms. One is the contribution of the local, same $\bR$ but both layers, terms, which, 
according to the results of the previous section, reads:
\be
E_{loc}= \Tr\,\left[\phi^\dagger\, \left(\mathcal{H}_U+\mathcal{H}_\perp-\mu\sum_{\bR i} n_{\bR,i}\right)\,\phi\right], 
\ee
where all operators are meant to be matrices in the local representation invariant under $SU(2)$ symmetry. 
The other contribution to the total energy is the intra-layer hopping $E_{hop}$. 
This can be shown to coincide with the ground-state energy of a variational 
single-particle Hamiltonian\cite{mio}:
\be\label{H-var}
\mathcal{H}_{hop}^*=\sum_\bk \psi^\dagger_\bk \, \hat{T}_\bk \, \psi_\bk ,
\ee
where $\psi^\dagger_\bk=(d^\dagger_{\bk, 1\up}, d^\dagger_{\bk, 2\up}, d^{\phantom{\dagger}}_{-\bk, 1\down}, 
d^{\phantom{\dagger}}_{-\bk, 2\down})$ is the Nambu spinor in momentum space and $\hat{T}_\bk$ a $4\times 4$ matrix in the natural basis
that depends explicitly on momentum and on some Lagrange multipliers. These are included to enforces that 
the average of the single particle density 
matrix on the ground state -- to be identified with $|\Psi_0\rangle$ 
in \eqn{GWF} -- is diagonal with matrix elements satisfying 
\[
\langle \Psi_0 | d^\dagger_{\bR i \sigma} d^\dagga_{\bR i \sigma} | \Psi_0\rangle = \Tr\left(\phi^\dagger\phi\,d^\dagger_{i\sigma}
d^\dagga_{i\sigma}\right) \equiv n^0_i.
\]
The matrix $\hat{T}_\bk$ has the general expression:
\be
\hat{T}_\bk = 
\left(
\begin{array}{cc}
 \epsilon(\bk)\hat{Z} +\hat{\eta}                  & \epsilon(\bk)\hat{\Delta} +\hat{\delta}    \\
 \epsilon(\bk)\hat{\Delta}^\dagger +\hat{\delta}^\dagger & -\epsilon(\bk)\hat{Z}^t -\hat{\eta}^t\\
\end{array}
\right),\label{H-var-explicit}
\ee
where the $2\times 2$ matrices $\hat{\eta}$ and $\hat{\delta}$ are the aforementioned Lagrange multipliers, while $\hat{Z}$ and 
$\hat{\Delta}$ have elements (labelled by $j,l=1,2$, the layer indices) 
\bea
\hat{Z}_{j,l} &=&\sum_{i=1}^2\,\left(R^\dagga_{i,j}\,R^*_{i,l}-Q^\dagga_{i,l}\,Q^*_{i,j}\right), \\
\hat{\Delta}_{j,l} &=&\sum_{i=1}^2\,\left(R^\dagga_{i,j}\,Q^*_{i,l}+R^\dagga_{i,l}\,Q^*_{i,j}\right). 
\eea

We solved numerically the variational problem assuming for simplicity a flat density of states with half-bandwidth $D$ (we
do not expect the results to change qualitatively by adopting a more realistic density of states). 
In order to compare with the half-filling results reported in Ref.~\onlinecite{mio}, we fixed the value of the intra-dimer hopping
$t_\perp/D=0.5$ and solved the variational problem for different values of $U/D$ and $\mu/D$. Note that this value in the case 
of a square lattice with nearest neighbor hopping $t$ corresponds to $t_\perp=2t$, above the critical value for 
the stability at large $U$ of the VB Mott insulator~\cite{J-Jperp-2}.  

\begin{figure}
\includegraphics[width=7cm]{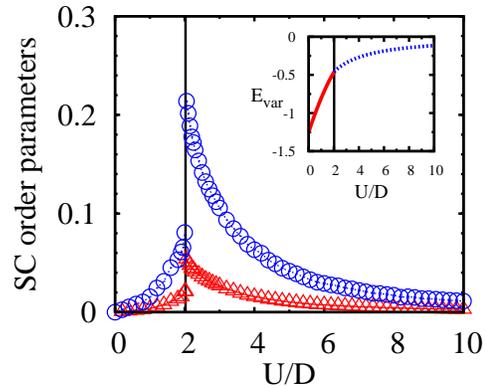}
\caption{(Color online) Inter-plane (blue circles) and, with reversed sign, intra-plane (red triangles)
superconducting order parameters at half-filling as function of $U/D$. The vertical line indicates the first order 
transition that we think identifies the on-set of Mott insulating behavior. Inset shows the variational energy in
units of $D$}\label{fig0}
\end{figure}

At half-filling, $\mu/D=0$ and we recover all results of Ref.~\onlinecite{mio}. Specifically, we find a first order 
metal to VB insulator transition. In the metallic phase just before the transition, singlet superconductivity
emerges. In Fig.~\ref{fig0} we show as function of $U/D$ the behavior of the inter-layer, $\Delta_\perp$ and 
intra-layer, $\Delta_{||}$, superconducting order parameters, defined as 
\bea
\Delta_\perp 
&=& \langle \Psi_G|\,c^\dagger_{\bR,1\up}c^\dagger_{\bR,2\down} + c^\dagger_{\bR,2\up}c^\dagger_{\bR,1\down}\,|\Psi_G\rangle,
\label{def-Delta*}\\
\Delta_{||} 
&=& \langle \Psi_G|\,c^\dagger_{\bR,i\up}c^\dagger_{\bRp,i\down} + c^\dagger_{\bRp,i\up}c^\dagger_{\bR,i\down}\,|\Psi_G\rangle,
\label{def-Delta_||}
\eea
where $\bR$ and $\bRp$ are nearest neighbor sites on layer $i=1,2$. We find that, near the first order transition 
that we think identifies the actual Mott transition, both order parameters are finite and have opposite sign, the 
so-called $d_{z^2-r^2}$ symmetry known to be dominant in the two-chain model~\cite{Scalapino-RVB}, and which QMC 
simulations~\cite{Bulut,dos-Santos} indicate as the leading pairing instability.  
The variational energy that we obtain appears to be slightly
lower than that found in Ref.~\onlinecite{mio}, as one could have expected due to the larger number of variational parameters. 
Nonetheless, the critical $U_c$ at the Mott transition is only slightly reduced to $U_c/D\simeq 2.02$ for $t_\perp/D=0.5$. 
We note that the phase at $U>U_c$, that we believe is Mott insulating, still shows a finite superconducting order parameter 
that dies out upon increasing $U$. As discussed in \onlinecite{mio}, we think this might be a spurious result of our variational 
approach that lacks intersite charge correlations crucial in stabilizing a genuine Mott insulating phase~\cite{Capello1}.

We study finite hole doping by varying $\mu/D<0$ at different values of $U/D$. 
Before discussing the variational results, we briefly sketch the behavior of the doped non-interacting system, $U/D=0$. The 
inter-layer coupling gives rise to bonding and antibonding bands, see Eq.~\eqn{Ham-e-o}. With the chosen value
of $t_\perp=0.5D$, these bands overlap at half-filling and the system displays a metallic behavior. 
When the chemical potential is lowered, 
holes are injected into the system inducing a depletion of both bands until, at a given value
of the chemical potential, the upper 
(antibonding) band empties. For the chosen $t_\perp$ and for a flat density of states the
complete depletion of the antibonding band happens at $\mu=0.5D$, corresponding to quarter filling $n=1$. As a consequence,
both the intra-layer ($E_{hop}$) and inter-layer ($E_\perp$) hopping contributions display a discontinuity in their first
derivatives at quarter filling, signaling that the antibonding band is no longer contributing. The
total energy however remains smooth for any value of $\mu$ (or equivalently $n$), as it should.
When $U/D\not =0$, the behavior that we find depends crucially if $U$ is smaller or greater than $U_c$, namely if the 
half-filled state is a metal or an insulator.

\begin{figure}
\includegraphics[width=7cm]{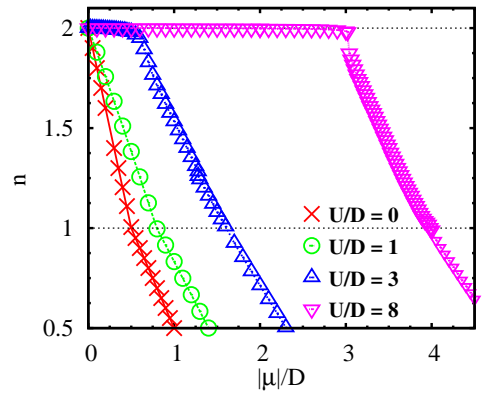}
\caption{(Color online) Average density $n$ summed over both layer as a function of the chemical potential $\mu<0$ 
for selected values of interaction
$U/D$.}\label{fig1}
\end{figure}

\begin{figure}[h]
\includegraphics[height=4.1cm]{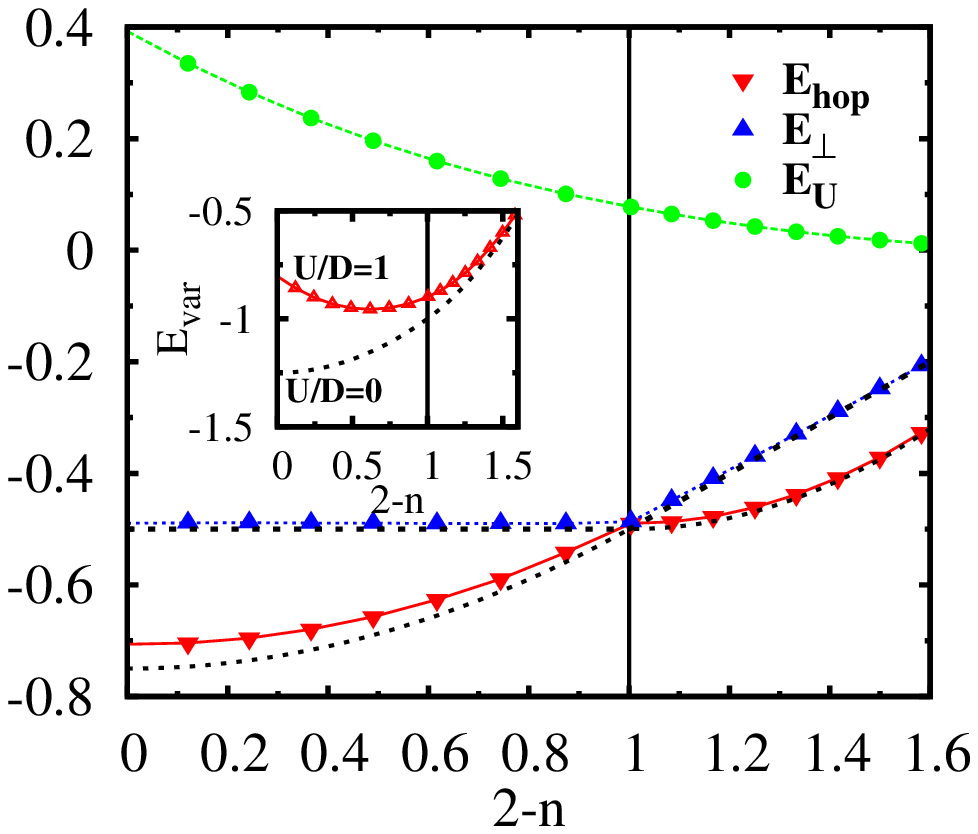}
\includegraphics[height=4.1cm]{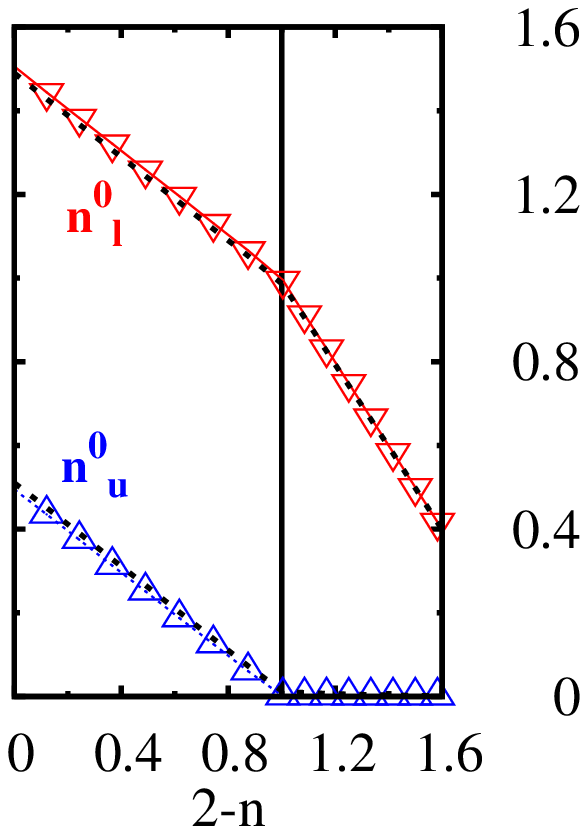}
\caption{(Color online) 
Left panel: The different contributions to the variational energy as a function of doping for $U/D=1$ and per lattice site $\bR$, i.e. 
summed over both layer. As a reference, the behavior of non-interacting inter- and intra-layer hopping contributions 
is plotted (dotted lines). In the inset the total energy $E_{var}(n)=E_{var}+\mu n$ is shown: 
despite the cusp observed in the hopping contributions, the
evolution of $E_{var}(n)$ is smooth. Right panel: Occupation of the variational lower and upper bands as function of $n$.
Dotted lines represent average occupation of even and odd orbitals.}
\label{fig3}
\end{figure}

\subsection{Doping the metal at $U<U_c$}

As long as $U< U_c$, any change of $\mu$ induces a continuous change in the total particle number; a finite compressibility signal 
of a metallic behavior, as shown in Fig.~\ref{fig1}. Alike the uncorrelated case, a cusp appears in the evolution
of $n$ at quarter-filling, that we explain seemingly as the depletion of the antibonding band.
Indeed, when $U<U_c$, the metallic solution evolves just like the non-interacting case. The main effect of interaction is to
slightly reduce inter- and intra-layer hopping contributions with respect to their uncorrelated counterparts,
as shown in Fig.~\ref{fig3} where we plot the
different contributions $E_{hop}$, $E_\perp$ and $E_U$ to the variational energy.
The intra-layer hopping contribution $E_{hop}$ diminishes in absolute value with increasing doping
because of the depletion of the bands, as it occurs in the non-interacting system; at quarter-filling it displays a cusp
and correspondingly the inter-layer hopping $E_\perp$ starts to rapidly decrease, the
effects of $U$ being more and more negligible as the low-density regime is approached. In the right panel of
Fig.~\ref{fig3} we show the occupancies $n^0_l$ and $n^0_u$ of the variational lower and upper bands, respectively, 
which are obtained by diagonalizing the associated variational Hamiltonian,
Eq.~(\ref{H-var}), and actually coincide with the eigenvalues of the single-particle density matrix. 
As in the uncorrelated system, the occupancy of the upper band vanishes at quarter filling. We stress the fact that
in the present approach these states are variationally determined and may not correspond to the even and odd combinations of
the original operators. However, as long as $U<U_c$, we find that the average values of bonding and antibonding band occupancies, $n_e$
and $n_o$, almost coincide with, respectively, $n^0_l$ and $n^0_u$.

Concerning superconductivity, we find that the inter-layer order parameter, Eq.~\eqn{def-Delta*}, 
is extremely small, practically zero within our numerical precision, see Fig.~\ref{fig2}. The intra-layer order parameter 
strictly follows the inter-layer one, hence is also zero.

\subsection{Doping the VB Mott insulator at $U>U_c$}

\begin{figure}[h]
\includegraphics[width=7cm]{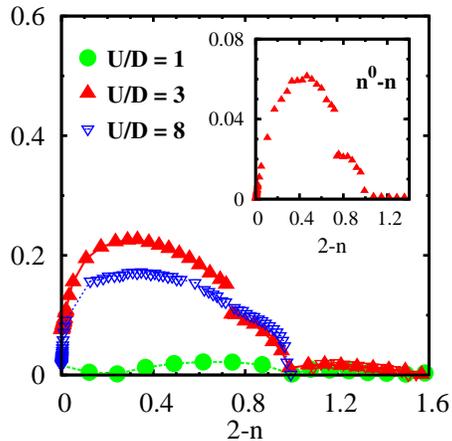}
\caption{(Color online) 
Superconducting inter-layer order parameter $\Delta_{\perp}$ for different $U/D$s. In the inset we plot the difference between the local
densities of the BCS variational wavefunction $|\Psi_0\rangle$ and of the actual one $|\Psi_G\rangle$, at 
$U/D=3$.}\label{fig2}
\end{figure}

When $U> U_c$, i.e. when the half-filled system is insulating,
the particle number remains stuck to its half-filled value $n=2$ until
$|\mu|\leq |\mu^*|\approx(U-U_c)/2$. This simply follows from the existence of the Mott gap at half-filling.
Upon doping, i.e. when $|\mu|>|\mu^*|$, a metallic behavior is clearly found. However, within our numerical
precision we can not establish whether the evolution from the insulator to the metal occurs smoothly (yet with a diverging 
compressibility) or through a weak first-order transition. Till the largest value of $U$ we considered, we could not find any
appreciable discontinuity in the evolution of $n$ at large doping, unlike for $U< U_c$ where a cusp is observed at quarter filling. 
In addition, contrary to the case $U<U_c$, here we find a clear superconducting signal between
half and quarter filling, see e.g. the behavior of $\Delta_{\perp}$, Eq.~\ref{def-Delta*}, 
shown in Fig.~\ref{fig2}. We note that $\Delta_{\perp}$ has a non-monotonous behavior, first increases quite rapidly with $U$ and 
for larger values decreases. Like at half-filling, a finite $\Delta_\perp$ produces through Eq.~\eqn{Z-qpnew} also 
a finite intra-layer $\Delta_{||}$, Eq.~\ref{def-Delta_||}, not shown here, which happens to have opposite sign.

\begin{figure}[h]
\includegraphics[height=4.1cm]{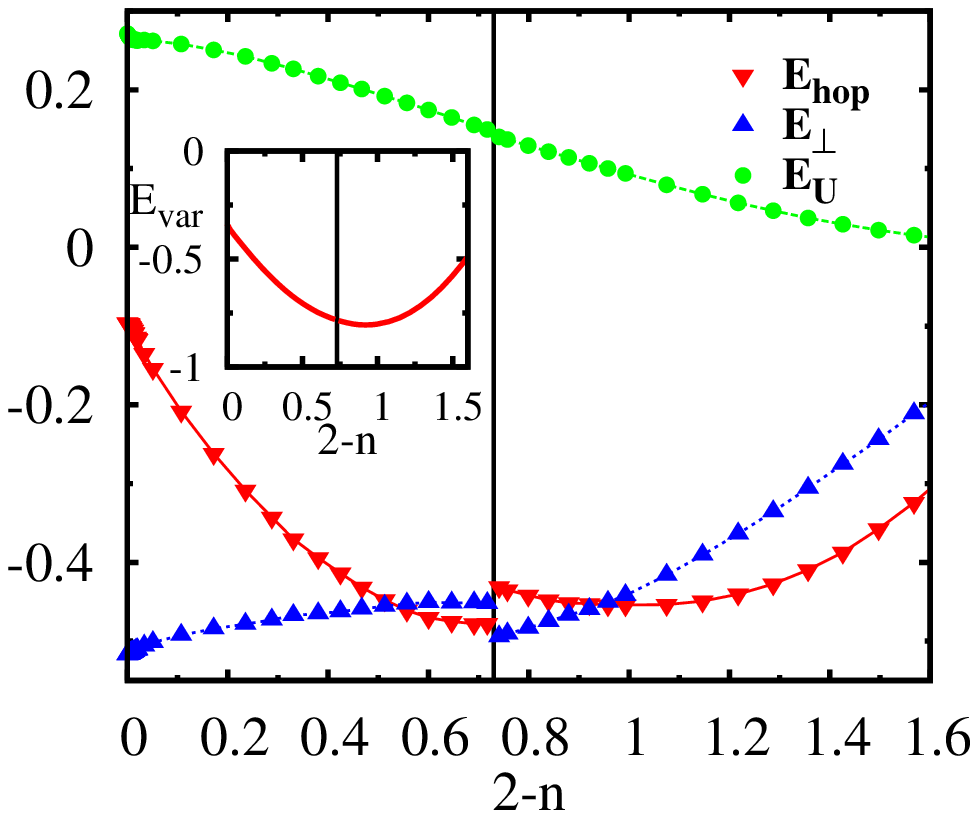}
\includegraphics[height=4.1cm]{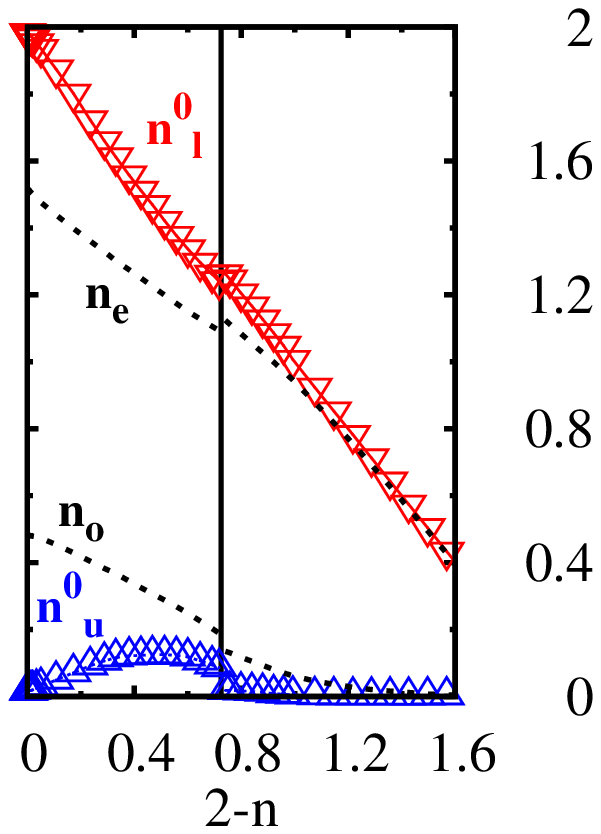}
\caption{(Color online) Left panel: Contributions to the variational energy as function of doping for $U/D=3$. In
the inset the total energy $E_{var}(n)=E_{var}+\mu n$ is shown: despite the discontinuities observed in the hopping contributions, the
evolution of $E(n)$ is, to our numerical accuracy, smooth. Right panel: Occupancies of lower, $n^0_l$ and upper,
$n^0_u$, variational bands 
as function of $n$. Dotted lines represent average occupation of even, $n_e$, and odd, $n_o$, orbitals. Note that the insulating phase 
at half-filling is identified by the lower band fully occupied and the upper one empty. The latter empties again for doping 
$2-n>0.73$.}
\label{fig4}
\end{figure}

Let us now consider in detail the energetic balance for $U>U_c$ and its differences with respect to $U<U_c$.  
At very large $U$ (not shown), as holes are injected into the system, both intra- and inter-layer hopping contributions 
first increase in absolute value, 
then saturate around approximatively quarter-filling, and eventually decrease as the low-density regime is attained, 
as expected when approaching the bottom of the variational bands. In other words, the behavior at large $U$ 
between half- and quarter-filling is quite different from the non-interacting case, while becomes quite similar below. 
This points to a very different influence 
of a strong interaction close to half-filling and far away from it and, indirectly, emphasizes the role of the 
superconductivity that we find for $2>n>1$.  For $U\gtrsim U_c$, i.e. closer to the half-filled metal-insulator transition, 
the picture is slightly different, as shown in Fig.~\ref{fig4} for $U/D=3$. 
To begin with, at small dopings the system gains in intra-layer hopping energy 
while the inter-layer one seems to be slightly reduced. Remarkably, even if the total energy is, within our numerical accuracy,  
a smooth function of $n$, both hopping contributions display a discontinuity at $\mu/D\simeq 1.28$, which corresponds to a local
density of $n\approx 1.27$. Here the occupation of the upper variational band goes to zero 
(cfr. right panel of Fig.~\ref{fig4}), even though 
nothing similar occurs in the occupation of the physical antibonding band. At this filling
fraction, the inter-layer hopping energy gain has an upward jump, contrary to the intra-layer one, even though further doping leads to 
a reduction of both.  A drop in the amplitude of the superconducting order parameter $\Delta_\perp$ 
is also found at this point. Further doping diminishes $\Delta_\perp$, which vanishes approximatively at quarter filling. 
A similar feature is observed in another quantity. Indeed, just like $n^0_l$ and $n^0_u$ may not correspond to the 
occupation of the bonding and antibonding bands, $n^0=n^0_l+n^0_u$, which is the average density of  
the BCS-like variational wavefunction, may differ from the physical one. In the inset of Fig.~\ref{fig2} we show
their difference for $U/D=3$. We observe that they actually deviate when superconductivity is found and their difference 
jumps down abruptly for $n< 1.27$.

\section{Conclusions}\label{Conclusions}

In this work we have studied by means of an extension of the Gutzwiller approximation the effect of doping a bilayer Hubbard model. 
We have considered a value of the inter-layer hopping $t_\perp$ such that, at half-filling, the model should undergo a 
direct transition at $U=U_c$ from a metal to a non-magnetic Mott insulator, a valence bond  
crystal consisting of inter-layer dimers. This choice offers the opportunity to study how a valence bond crystal liquefies either 
by reducing the Coulomb repulsion keeping the 
density fixed at one electron per site, or by adding mobile holes. The melting upon decreasing $U$ was already shown~\cite{mio} to 
lead to a superconducting phase intruding between the valence bond insulator at large $U>U_c$ and the normal metal at weak $U\ll U_c$. 
Here we show that superconductivity arises also upon melting the valence bond crystal by doping. 
In other words, the superconducting dome 
that exists at half-filling close to $U_c$ extends into a whole region at finite doping. The maximum superconducting signal 
is found at 20\% doping, and beyond that it smoothly diminishes, disappearing roughly at quarter filling within our choice of 
parameters. These results are appealing as they show that the well established behavior of a two-leg Hubbard 
ladder~\cite{mio,Balents&Fisher,Schulz,Scalapino-RVB} seems to survive in higher dimensions, actually in the infinite-dimension 
limit where our Gutzwiller approximation becomes exact. It is obvious that, in spite of all improvements of the Gutzwiller 
variational approach, to which we contribute a bit with this work, this method remains variational hence not exact. Therefore 
it is still under question if superconductivity indeed arises by metallizing the valence bond Mott 
insulating phase of a Hubbard bilayer, which we believe is an important issue of broader interest than the simple 
bilayer model we have investigated~\cite{Capone04}.
There are actually quantum Monte Carlo simulations~\cite{Scalettar,Bulut,dos-Santos,Referee} that partially support our 
results as they show a pronounced enhancement of superconducting fluctuations close to the half-filled Mott insulator. 
However a true superconducting phase is still unaccessible to the lowest temperatures that can be reached by quantum 
Monte Carlo. On the other hand, dynamical mean field calculations, that can access zero temperature phases, did not so far 
looked for superconductivity~\cite{Monien,Okamoto}. Therefore we think it would be worth pursuing further this issue.


\begin{thebibliography}{50}
\expandafter\ifx\csname natexlab\endcsname\relax\def\natexlab#1{#1}\fi
\expandafter\ifx\csname bibnamefont\endcsname\relax
  \def\bibnamefont#1{#1}\fi
\expandafter\ifx\csname bibfnamefont\endcsname\relax
  \def\bibfnamefont#1{#1}\fi
\expandafter\ifx\csname citenamefont\endcsname\relax
  \def\citenamefont#1{#1}\fi
\expandafter\ifx\csname url\endcsname\relax
  \def\url#1{\texttt{#1}}\fi
\expandafter\ifx\csname urlprefix\endcsname\relax\def\urlprefix{URL }\fi
\providecommand{\bibinfo}[2]{#2}
\providecommand{\eprint}[2][]{\url{#2}}

\bibitem[{\citenamefont{Gutzwiller}(1963)}]{Gutzwiller1}
\bibinfo{author}{\bibfnamefont{M.~C.} \bibnamefont{Gutzwiller}},
  \bibinfo{journal}{Phys. Rev. Lett.} \textbf{\bibinfo{volume}{10}},
  \bibinfo{pages}{159} (\bibinfo{year}{1963}).

\bibitem[{\citenamefont{Gutzwiller}(1964)}]{Gutzwiller2}
\bibinfo{author}{\bibfnamefont{M.~C.} \bibnamefont{Gutzwiller}},
  \bibinfo{journal}{Phys. Rev.} \textbf{\bibinfo{volume}{134}},
  \bibinfo{pages}{A923} (\bibinfo{year}{1964}).

\bibitem[{\citenamefont{Gutzwiller}(1965)}]{Gutzwiller3}
\bibinfo{author}{\bibfnamefont{M.~C.} \bibnamefont{Gutzwiller}},
  \bibinfo{journal}{Phys. Rev.} \textbf{\bibinfo{volume}{137}},
  \bibinfo{pages}{A1726} (\bibinfo{year}{1965}).

\bibitem[{\citenamefont{Sorella}(2005)}]{Sorella-VMC}
\bibinfo{author}{\bibfnamefont{S.}~\bibnamefont{Sorella}},
  \bibinfo{journal}{Phys. Rev. B} \textbf{\bibinfo{volume}{71}},
  \bibinfo{pages}{241103} (\bibinfo{year}{2005}).

\bibitem[{\citenamefont{M{\"{u}ller}-Hartmann}(1989)}]{Muller}
\bibinfo{author}{\bibfnamefont{E.}~\bibnamefont{M{\"{u}ller}-Hartmann}},
  \bibinfo{journal}{Z. Phys. B: Condens. Matter} \textbf{\bibinfo{volume}{76}},
  \bibinfo{pages}{211} (\bibinfo{year}{1989}).

\bibitem[{\citenamefont{Metzner and Vollhardt}(1987)}]{Metzner-Vollhardt-PRL}
\bibinfo{author}{\bibfnamefont{W.}~\bibnamefont{Metzner}} \bibnamefont{and}
  \bibinfo{author}{\bibfnamefont{D.}~\bibnamefont{Vollhardt}},
  \bibinfo{journal}{Phys. Rev. Lett.} \textbf{\bibinfo{volume}{59}},
  \bibinfo{pages}{121} (\bibinfo{year}{1987}).

\bibitem[{\citenamefont{Metzner and Vollhardt}(1988)}]{Metzner-Vollhardt-PRB}
\bibinfo{author}{\bibfnamefont{W.}~\bibnamefont{Metzner}} \bibnamefont{and}
  \bibinfo{author}{\bibfnamefont{D.}~\bibnamefont{Vollhardt}},
  \bibinfo{journal}{Phys. Rev. B} \textbf{\bibinfo{volume}{37}},
  \bibinfo{pages}{7382} (\bibinfo{year}{1988}).

\bibitem[{\citenamefont{B\"unemann et~al.}(1998)\citenamefont{B\"unemann,
  Weber, and Gebhard}}]{Gebhard}
\bibinfo{author}{\bibfnamefont{J.}~\bibnamefont{B\"unemann}},
  \bibinfo{author}{\bibfnamefont{W.}~\bibnamefont{Weber}}, \bibnamefont{and}
  \bibinfo{author}{\bibfnamefont{F.}~\bibnamefont{Gebhard}},
  \bibinfo{journal}{Phys. Rev. B} \textbf{\bibinfo{volume}{57}},
  \bibinfo{pages}{6896} (\bibinfo{year}{1998}).

\bibitem[{\citenamefont{Kotliar and Ruckenstein}(1986)}]{Kotliar-Ruckenstein}
\bibinfo{author}{\bibfnamefont{G.}~\bibnamefont{Kotliar}} \bibnamefont{and}
  \bibinfo{author}{\bibfnamefont{A.~E.} \bibnamefont{Ruckenstein}},
  \bibinfo{journal}{Phys. Rev. Lett.} \textbf{\bibinfo{volume}{57}},
  \bibinfo{pages}{1362} (\bibinfo{year}{1986}).

\bibitem[{\citenamefont{Bunemann and Gebhard}(2007)}]{quaquaraqua}
\bibinfo{author}{\bibfnamefont{J.}~\bibnamefont{Bunemann}} \bibnamefont{and}
  \bibinfo{author}{\bibfnamefont{F.}~\bibnamefont{Gebhard}},
  \bibinfo{journal}{Phys. Rev.B} \textbf{\bibinfo{volume}{76}},
  \bibinfo{eid}{193104} (\bibinfo{year}{2007}).

\bibitem[{\citenamefont{Lechermann et~al.}(2007)\citenamefont{Lechermann,
  Georges, Kotliar, and Parcollet}}]{Georges}
\bibinfo{author}{\bibfnamefont{F.}~\bibnamefont{Lechermann}},
  \bibinfo{author}{\bibfnamefont{A.}~\bibnamefont{Georges}},
  \bibinfo{author}{\bibfnamefont{G.}~\bibnamefont{Kotliar}}, \bibnamefont{and}
  \bibinfo{author}{\bibfnamefont{O.}~\bibnamefont{Parcollet}},
  \bibinfo{journal}{Phys. Rev. B} \textbf{\bibinfo{volume}{76}},
  \bibinfo{eid}{155102} (\bibinfo{year}{2007}).

\bibitem[{\citenamefont{Brinkman and Rice}(1970)}]{Brinkman&Rice}
\bibinfo{author}{\bibfnamefont{W.~F.} \bibnamefont{Brinkman}} \bibnamefont{and}
  \bibinfo{author}{\bibfnamefont{T.~M.} \bibnamefont{Rice}},
  \bibinfo{journal}{Phys. Rev. B} \textbf{\bibinfo{volume}{2}},
  \bibinfo{pages}{4302} (\bibinfo{year}{1970}).

\bibitem[{\citenamefont{Georges et~al.}(1996)\citenamefont{Georges, Kotliar,
  Krauth, and Rozenberg}}]{DMFT}
\bibinfo{author}{\bibfnamefont{A.}~\bibnamefont{Georges}},
  \bibinfo{author}{\bibfnamefont{G.}~\bibnamefont{Kotliar}},
  \bibinfo{author}{\bibfnamefont{W.}~\bibnamefont{Krauth}}, \bibnamefont{and}
  \bibinfo{author}{\bibfnamefont{M.~J.} \bibnamefont{Rozenberg}},
  \bibinfo{journal}{Rev. Mod. Phys.} \textbf{\bibinfo{volume}{68}},
  \bibinfo{pages}{13} (\bibinfo{year}{1996}).

\bibitem[{\citenamefont{Anisimov et~al.}(1997)\citenamefont{Anisimov,
  Aryasetiawan, and Lichtenstein}}]{LDA+U}
\bibinfo{author}{\bibfnamefont{V.~I.} \bibnamefont{Anisimov}},
  \bibinfo{author}{\bibfnamefont{F.}~\bibnamefont{Aryasetiawan}},
  \bibnamefont{and}
  \bibinfo{author}{\bibfnamefont{A.}~\bibnamefont{Lichtenstein}},
  \bibinfo{journal}{J. Phys.: Condens. Matter} \textbf{\bibinfo{volume}{9}},
  \bibinfo{pages}{767} (\bibinfo{year}{1997}).

\bibitem[{\citenamefont{Capello et~al.}(2005)\citenamefont{Capello, Becca,
  Fabrizio, Sorella, and Tosatti}}]{Capello1}
\bibinfo{author}{\bibfnamefont{M.}~\bibnamefont{Capello}},
  \bibinfo{author}{\bibfnamefont{F.}~\bibnamefont{Becca}},
  \bibinfo{author}{\bibfnamefont{M.}~\bibnamefont{Fabrizio}},
  \bibinfo{author}{\bibfnamefont{S.}~\bibnamefont{Sorella}}, \bibnamefont{and}
  \bibinfo{author}{\bibfnamefont{E.}~\bibnamefont{Tosatti}},
  \bibinfo{journal}{Phys. Rev. Lett.} \textbf{\bibinfo{volume}{94}},
  \bibinfo{pages}{026406} (\bibinfo{year}{2005}).

\bibitem[{\citenamefont{B\"unemann et~al.}(2005)\citenamefont{B\"unemann,
  Gebhard, Ohm, Weiser, and Weber}}]{Bunemann}
\bibinfo{author}{\bibfnamefont{J.}~\bibnamefont{B\"unemann}},
  \bibinfo{author}{\bibfnamefont{F.}~\bibnamefont{Gebhard}},
  \bibinfo{author}{\bibfnamefont{T.}~\bibnamefont{Ohm}},
  \bibinfo{author}{\bibfnamefont{S.}~\bibnamefont{Weiser}}, \bibnamefont{and}
  \bibinfo{author}{\bibfnamefont{W.}~\bibnamefont{Weber}}, in
  \emph{\bibinfo{booktitle}{Frontiers in Magnetic Materials}}, edited by
  \bibinfo{editor}{\bibfnamefont{A.}~\bibnamefont{Narlikar}}
  (\bibinfo{publisher}{Springer, Berlin}, \bibinfo{year}{2005}), pp.
  \bibinfo{pages}{117--151}.

\bibitem[{\citenamefont{Attaccalite and Fabrizio}(2003)}]{Attaccalite}
\bibinfo{author}{\bibfnamefont{C.}~\bibnamefont{Attaccalite}} \bibnamefont{and}
  \bibinfo{author}{\bibfnamefont{M.}~\bibnamefont{Fabrizio}},
  \bibinfo{journal}{Phys. Rev. B} \textbf{\bibinfo{volume}{68}},
  \bibinfo{pages}{155117} (\bibinfo{year}{2003}).

\bibitem[{\citenamefont{Fabrizio}(2007)}]{mio}
\bibinfo{author}{\bibfnamefont{M.}~\bibnamefont{Fabrizio}},
  \bibinfo{journal}{Phys. Rev. B} \textbf{\bibinfo{volume}{76}},
  \bibinfo{eid}{165110} (\bibinfo{year}{2007}).

\bibitem[{\citenamefont{Barone et~al.}(2007)\citenamefont{Barone, Raimondi,
  Capone, Castellani, and Fabrizio}}]{Barone1}
\bibinfo{author}{\bibfnamefont{P.}~\bibnamefont{Barone}},
  \bibinfo{author}{\bibfnamefont{R.}~\bibnamefont{Raimondi}},
  \bibinfo{author}{\bibfnamefont{M.}~\bibnamefont{Capone}},
  \bibinfo{author}{\bibfnamefont{C.}~\bibnamefont{Castellani}},
  \bibnamefont{and} \bibinfo{author}{\bibfnamefont{M.}~\bibnamefont{Fabrizio}},
  \bibinfo{journal}{Europhys. Lett.} \textbf{\bibinfo{volume}{79}},
  \bibinfo{pages}{47003} (\bibinfo{year}{2007}).

\bibitem[{\citenamefont{Borghi et~al.}(2009)\citenamefont{Borghi, Fabrizio, and
  Tosatti}}]{Borghi}
\bibinfo{author}{\bibfnamefont{G.}~\bibnamefont{Borghi}},
  \bibinfo{author}{\bibfnamefont{M.}~\bibnamefont{Fabrizio}}, \bibnamefont{and}
  \bibinfo{author}{\bibfnamefont{E.}~\bibnamefont{Tosatti}},
  \bibinfo{journal}{Phys. Rev. Lett.} \textbf{\bibinfo{volume}{102}},
  \bibinfo{eid}{066806}  (\bibinfo{year}{2009}).

\bibitem[{\citenamefont{Deng et~al.}(2009)\citenamefont{Deng, Wang, Dai, and
  Fang}}]{Fang}
\bibinfo{author}{\bibfnamefont{X.}~\bibnamefont{Deng}},
  \bibinfo{author}{\bibfnamefont{L.}~\bibnamefont{Wang}},
  \bibinfo{author}{\bibfnamefont{X.}~\bibnamefont{Dai}}, \bibnamefont{and}
  \bibinfo{author}{\bibfnamefont{Z.}~\bibnamefont{Fang}},
  \bibinfo{journal}{Phys. Rev. B} \textbf{\bibinfo{volume}{79}},
  \bibinfo{eid}{075114}  (\bibinfo{year}{2009}).

\bibitem[{\citenamefont{Bickers et~al.}(1989)\citenamefont{Bickers, Scalapino,
  and White}}]{Scalapino}
\bibinfo{author}{\bibfnamefont{N.~E.} \bibnamefont{Bickers}},
  \bibinfo{author}{\bibfnamefont{D.~J.} \bibnamefont{Scalapino}},
  \bibnamefont{and} \bibinfo{author}{\bibfnamefont{S.~R.} \bibnamefont{White}},
  \bibinfo{journal}{Phys. Rev. Lett.} \textbf{\bibinfo{volume}{62}},
  \bibinfo{pages}{961} (\bibinfo{year}{1989}).

\bibitem[{\citenamefont{Kotliar and Liu}(1988)}]{Kotliar&Liu}
\bibinfo{author}{\bibfnamefont{G.}~\bibnamefont{Kotliar}} \bibnamefont{and}
  \bibinfo{author}{\bibfnamefont{J.}~\bibnamefont{Liu}},
  \bibinfo{journal}{Phys. Rev. B} \textbf{\bibinfo{volume}{38}},
  \bibinfo{pages}{5142} (\bibinfo{year}{1988}).

\bibitem[{\citenamefont{Gros}(1988)}]{Gros-PRB}
\bibinfo{author}{\bibfnamefont{C.}~\bibnamefont{Gros}}, \bibinfo{journal}{Phys.
  Rev. B} \textbf{\bibinfo{volume}{38}}, \bibinfo{pages}{931}
  (\bibinfo{year}{1988}).

\bibitem[{\citenamefont{Paramekanti et~al.}(2004)\citenamefont{Paramekanti,
  Randeria, and Trivedi}}]{Paramekanti}
\bibinfo{author}{\bibfnamefont{A.}~\bibnamefont{Paramekanti}},
  \bibinfo{author}{\bibfnamefont{M.}~\bibnamefont{Randeria}}, \bibnamefont{and}
  \bibinfo{author}{\bibfnamefont{N.}~\bibnamefont{Trivedi}},
  \bibinfo{journal}{Phys. Rev. B} \textbf{\bibinfo{volume}{70}},
  \bibinfo{pages}{054504} (\bibinfo{year}{2004}).

\bibitem[{\citenamefont{Sorella et~al.}(2002)\citenamefont{Sorella, Martins,
  Becca, Gazza, Capriotti, Parola, and Dagotto}}]{Sandro&Dagotto}
\bibinfo{author}{\bibfnamefont{S.}~\bibnamefont{Sorella}},
  \bibinfo{author}{\bibfnamefont{G.~B.} \bibnamefont{Martins}},
  \bibinfo{author}{\bibfnamefont{F.}~\bibnamefont{Becca}},
  \bibinfo{author}{\bibfnamefont{C.}~\bibnamefont{Gazza}},
  \bibinfo{author}{\bibfnamefont{L.}~\bibnamefont{Capriotti}},
  \bibinfo{author}{\bibfnamefont{A.}~\bibnamefont{Parola}}, \bibnamefont{and}
  \bibinfo{author}{\bibfnamefont{E.}~\bibnamefont{Dagotto}},
  \bibinfo{journal}{Phys. Rev. Lett.} \textbf{\bibinfo{volume}{88}},
  \bibinfo{pages}{117002} (\bibinfo{year}{2002}).

\bibitem[{\citenamefont{Anderson et~al.}(2004)\citenamefont{Anderson, Lee,
  Randeria, Rice, Trivedi, and Zhang}}]{Vanilla}
\bibinfo{author}{\bibfnamefont{P.~W.} \bibnamefont{Anderson}},
  \bibinfo{author}{\bibfnamefont{P.~A.} \bibnamefont{Lee}},
  \bibinfo{author}{\bibfnamefont{M.}~\bibnamefont{Randeria}},
  \bibinfo{author}{\bibfnamefont{T.~M.} \bibnamefont{Rice}},
  \bibinfo{author}{\bibfnamefont{N.}~\bibnamefont{Trivedi}}, \bibnamefont{and}
  \bibinfo{author}{\bibfnamefont{F.~C.} \bibnamefont{Zhang}},
  \bibinfo{journal}{J. Phys.: Condens. Matter} \textbf{\bibinfo{volume}{16}},
  \bibinfo{pages}{R755} (\bibinfo{year}{2004}).

\bibitem[{\citenamefont{Lichtenstein and
  Katsnelson}(2000)}]{CDMFT-Lichtenstein}
\bibinfo{author}{\bibfnamefont{A.~I.} \bibnamefont{Lichtenstein}}
  \bibnamefont{and} \bibinfo{author}{\bibfnamefont{M.~I.}
  \bibnamefont{Katsnelson}}, \bibinfo{journal}{Phys. Rev. B}
  \textbf{\bibinfo{volume}{62}}, \bibinfo{pages}{R9283} (\bibinfo{year}{2000}).

\bibitem[{\citenamefont{Potthoff et~al.}(2003)\citenamefont{Potthoff, Aichhorn,
  and Dahnken}}]{CDMFT-Potthoff}
\bibinfo{author}{\bibfnamefont{M.}~\bibnamefont{Potthoff}},
  \bibinfo{author}{\bibfnamefont{M.}~\bibnamefont{Aichhorn}}, \bibnamefont{and}
  \bibinfo{author}{\bibfnamefont{C.}~\bibnamefont{Dahnken}},
  \bibinfo{journal}{Phys. Rev. Lett.} \textbf{\bibinfo{volume}{91}},
  \bibinfo{pages}{206402} (\bibinfo{year}{2003}).

\bibitem[{\citenamefont{Senechal et~al.}(2000)\citenamefont{Senechal, Perez,
  and Pioro-Ladriere}}]{CDMFT-Senechal}
\bibinfo{author}{\bibfnamefont{D.}~\bibnamefont{Senechal}},
  \bibinfo{author}{\bibfnamefont{D.}~\bibnamefont{Perez}}, \bibnamefont{and}
  \bibinfo{author}{\bibfnamefont{M.}~\bibnamefont{Pioro-Ladriere}},
  \bibinfo{journal}{Phys. Rev. Lett.} \textbf{\bibinfo{volume}{84}},
  \bibinfo{pages}{522} (\bibinfo{year}{2000}).

\bibitem[{\citenamefont{Kotliar et~al.}(2001)\citenamefont{Kotliar, Savrasov,
  P\'alsson, and Biroli}}]{CDMFT-Kotliar}
\bibinfo{author}{\bibfnamefont{G.}~\bibnamefont{Kotliar}},
  \bibinfo{author}{\bibfnamefont{S.~Y.} \bibnamefont{Savrasov}},
  \bibinfo{author}{\bibfnamefont{G.}~\bibnamefont{P\'alsson}},
  \bibnamefont{and} \bibinfo{author}{\bibfnamefont{G.}~\bibnamefont{Biroli}},
  \bibinfo{journal}{Phys. Rev. Lett.} \textbf{\bibinfo{volume}{87}},
  \bibinfo{pages}{186401} (\bibinfo{year}{2001}).

\bibitem[{\citenamefont{Maier et~al.}(2005)\citenamefont{Maier, Jarrell,
  Pruschke, and Hettler}}]{CDMFT-Jarrell}
\bibinfo{author}{\bibfnamefont{T.}~\bibnamefont{Maier}},
  \bibinfo{author}{\bibfnamefont{M.}~\bibnamefont{Jarrell}},
  \bibinfo{author}{\bibfnamefont{T.}~\bibnamefont{Pruschke}}, \bibnamefont{and}
  \bibinfo{author}{\bibfnamefont{M.~H.} \bibnamefont{Hettler}},
  \bibinfo{journal}{Rev. Mod. Phys.} \textbf{\bibinfo{volume}{77}},
  \bibinfo{pages}{1027} (\bibinfo{year}{2005}).

\bibitem[{\citenamefont{Dahnken et~al.}(2004)\citenamefont{Dahnken, Aichhorn,
  Hanke, Arrigoni, and Potthoff}}]{Arrigoni}
\bibinfo{author}{\bibfnamefont{C.}~\bibnamefont{Dahnken}},
  \bibinfo{author}{\bibfnamefont{M.}~\bibnamefont{Aichhorn}},
  \bibinfo{author}{\bibfnamefont{W.}~\bibnamefont{Hanke}},
  \bibinfo{author}{\bibfnamefont{E.}~\bibnamefont{Arrigoni}}, \bibnamefont{and}
  \bibinfo{author}{\bibfnamefont{M.}~\bibnamefont{Potthoff}},
  \bibinfo{journal}{Phys. Rev. B} \textbf{\bibinfo{volume}{70}},
  \bibinfo{pages}{245110} (\bibinfo{year}{2004}).

\bibitem[{\citenamefont{Strong and Millis}(1994)}]{Strong&Millis}
\bibinfo{author}{\bibfnamefont{S.~P.} \bibnamefont{Strong}} \bibnamefont{and}
  \bibinfo{author}{\bibfnamefont{A.~J.} \bibnamefont{Millis}},
  \bibinfo{journal}{Phys. Rev. B} \textbf{\bibinfo{volume}{50}},
  \bibinfo{pages}{9911} (\bibinfo{year}{1994}).

\bibitem[{\citenamefont{Shelton et~al.}(1996)\citenamefont{Shelton, Nersesyan,
  and Tsvelik}}]{shura}
\bibinfo{author}{\bibfnamefont{D.~G.} \bibnamefont{Shelton}},
  \bibinfo{author}{\bibfnamefont{A.~A.} \bibnamefont{Nersesyan}},
  \bibnamefont{and} \bibinfo{author}{\bibfnamefont{A.~M.}
  \bibnamefont{Tsvelik}}, \bibinfo{journal}{Phys. Rev. B}
  \textbf{\bibinfo{volume}{53}}, \bibinfo{pages}{8521} (\bibinfo{year}{1996}).

\bibitem[{\citenamefont{Balents and Fisher}(1996)}]{Balents&Fisher}
\bibinfo{author}{\bibfnamefont{L.}~\bibnamefont{Balents}} \bibnamefont{and}
  \bibinfo{author}{\bibfnamefont{M.~P.~A.} \bibnamefont{Fisher}},
  \bibinfo{journal}{Phys. Rev. B} \textbf{\bibinfo{volume}{53}},
  \bibinfo{pages}{12133} (\bibinfo{year}{1996}).

\bibitem[{\citenamefont{Sierra et~al.}(1998)\citenamefont{Sierra,
  Mart\'in-Delgado, Dukelsky, White, and Scalapino}}]{Scalapino-RVB}
\bibinfo{author}{\bibfnamefont{G.}~\bibnamefont{Sierra}},
  \bibinfo{author}{\bibfnamefont{M.~A.} \bibnamefont{Mart\'in-Delgado}},
  \bibinfo{author}{\bibfnamefont{J.}~\bibnamefont{Dukelsky}},
  \bibinfo{author}{\bibfnamefont{S.~R.} \bibnamefont{White}}, \bibnamefont{and}
  \bibinfo{author}{\bibfnamefont{D.~J.} \bibnamefont{Scalapino}},
  \bibinfo{journal}{Phys. Rev. B} \textbf{\bibinfo{volume}{57}},
  \bibinfo{pages}{11666} (\bibinfo{year}{1998}).

\bibitem[{\citenamefont{Fabrizio}(1993)}]{2chain}
\bibinfo{author}{\bibfnamefont{M.}~\bibnamefont{Fabrizio}},
  \bibinfo{journal}{Phys. Rev. B} \textbf{\bibinfo{volume}{48}},
  \bibinfo{pages}{15838} (\bibinfo{year}{1993}).

\bibitem[{\citenamefont{Schulz}(1999)}]{Schulz}
\bibinfo{author}{\bibfnamefont{H.~J.} \bibnamefont{Schulz}},
  \bibinfo{journal}{Phys. Rev. B} \textbf{\bibinfo{volume}{59}},
  \bibinfo{pages}{R2471} (\bibinfo{year}{1999}).

\bibitem[{\citenamefont{Anderson}(1987)}]{PWA}
\bibinfo{author}{\bibfnamefont{P.~W.} \bibnamefont{Anderson}},
  \bibinfo{journal}{Science} \textbf{\bibinfo{volume}{235}},
  \bibinfo{pages}{1196} (\bibinfo{year}{1987}).

\bibitem[{\citenamefont{Sandvik and Scalapino}(1994)}]{J-Jperp-1}
\bibinfo{author}{\bibfnamefont{A.~W.} \bibnamefont{Sandvik}} \bibnamefont{and}
  \bibinfo{author}{\bibfnamefont{D.~J.} \bibnamefont{Scalapino}},
  \bibinfo{journal}{Phys. Rev. Lett.} \textbf{\bibinfo{volume}{72}},
  \bibinfo{pages}{2777} (\bibinfo{year}{1994}).

\bibitem[{\citenamefont{Scalettar et~al.}(1994)\citenamefont{Scalettar, Cannon,
  Scalapino, and Sugar}}]{Scalettar}
\bibinfo{author}{\bibfnamefont{R.~T.} \bibnamefont{Scalettar}},
  \bibinfo{author}{\bibfnamefont{J.~W.} \bibnamefont{Cannon}},
  \bibinfo{author}{\bibfnamefont{D.~J.} \bibnamefont{Scalapino}},
  \bibnamefont{and} \bibinfo{author}{\bibfnamefont{R.~L.} \bibnamefont{Sugar}},
  \bibinfo{journal}{Phys. Rev. B} \textbf{\bibinfo{volume}{50}},
  \bibinfo{pages}{13419} (\bibinfo{year}{1994}).

\bibitem[{\citenamefont{dos Santos}(1995)}]{dos-Santos}
\bibinfo{author}{\bibfnamefont{R.~R.} \bibnamefont{dos Santos}},
  \bibinfo{journal}{Phys. Rev. B} \textbf{\bibinfo{volume}{51}},
  \bibinfo{pages}{15540} (\bibinfo{year}{1995}).

\bibitem[{\citenamefont{Wang et~al.}(2006)\citenamefont{Wang, Beach, and
  Sandvik}}]{J-Jperp-2}
\bibinfo{author}{\bibfnamefont{L.}~\bibnamefont{Wang}},
  \bibinfo{author}{\bibfnamefont{K.~S.~D.} \bibnamefont{Beach}},
  \bibnamefont{and} \bibinfo{author}{\bibfnamefont{A.~W.}
  \bibnamefont{Sandvik}}, \bibinfo{journal}{Phys. Rev. B}
  \textbf{\bibinfo{volume}{73}}, \bibinfo{eid}{014431} (\bibinfo{year}{2006}).

\bibitem[{\citenamefont{Bouadim et~al.}(2008)\citenamefont{Bouadim, Batrouni,
  H\'{e}bert, and Scalettar}}]{Referee}
\bibinfo{author}{\bibfnamefont{K.}~\bibnamefont{Bouadim}},
  \bibinfo{author}{\bibfnamefont{G.~G.} \bibnamefont{Batrouni}},
  \bibinfo{author}{\bibfnamefont{F.}~\bibnamefont{H\'{e}bert}},
  \bibnamefont{and} \bibinfo{author}{\bibfnamefont{R.~T.}
  \bibnamefont{Scalettar}}, \bibinfo{journal}{Phys. Rev. B} \textbf{\bibinfo{volume}{77}},
  \bibinfo{eid}{144527}  (\bibinfo{year}{2008}).

\bibitem[{\citenamefont{Fuhrmann et~al.}(2006)\citenamefont{Fuhrmann, Heilmann,
  and Monien}}]{Monien}
\bibinfo{author}{\bibfnamefont{A.}~\bibnamefont{Fuhrmann}},
  \bibinfo{author}{\bibfnamefont{D.}~\bibnamefont{Heilmann}}, \bibnamefont{and}
  \bibinfo{author}{\bibfnamefont{H.}~\bibnamefont{Monien}},
  \bibinfo{journal}{Phys. Rev. B} \textbf{\bibinfo{volume}{73}},
  \bibinfo{pages}{245118} (\bibinfo{year}{2006}).

\bibitem[{\citenamefont{Kancharla and Okamoto}(2007)}]{Okamoto}
\bibinfo{author}{\bibfnamefont{S.~S.} \bibnamefont{Kancharla}}
  \bibnamefont{and} \bibinfo{author}{\bibfnamefont{S.}~\bibnamefont{Okamoto}},
  \bibinfo{journal}{Phys. Rev. B} \textbf{\bibinfo{volume}{75}},
  \bibinfo{eid}{193103} (\bibinfo{year}{2007}).

\bibitem[{\citenamefont{Bulut et~al.}(1992)\citenamefont{Bulut, Scalapino, and
  Scalettar}}]{Bulut}
\bibinfo{author}{\bibfnamefont{N.}~\bibnamefont{Bulut}},
  \bibinfo{author}{\bibfnamefont{D.~J.} \bibnamefont{Scalapino}},
  \bibnamefont{and} \bibinfo{author}{\bibfnamefont{R.~T.}
  \bibnamefont{Scalettar}}, \bibinfo{journal}{Phys. Rev. B}
  \textbf{\bibinfo{volume}{45}}, \bibinfo{pages}{5577} (\bibinfo{year}{1992}).

\bibitem[{\citenamefont{Lanat\`a et~al.}(2008)\citenamefont{Lanat\`a, Barone,
  and Fabrizio}}]{nostro}
\bibinfo{author}{\bibfnamefont{N.}~\bibnamefont{Lanat\`a}},
  \bibinfo{author}{\bibfnamefont{P.}~\bibnamefont{Barone}}, \bibnamefont{and}
  \bibinfo{author}{\bibfnamefont{M.}~\bibnamefont{Fabrizio}},
  \bibinfo{journal}{Phys. Rev. B} \textbf{\bibinfo{volume}{78}},
  \bibinfo{pages}{155127} (\bibinfo{year}{2008}).

\bibitem[{\citenamefont{Capone et~al.}(2004)\citenamefont{Capone, Fabrizio,
  Castellani, and Tosatti}}]{Capone04}
\bibinfo{author}{\bibfnamefont{M.}~\bibnamefont{Capone}},
  \bibinfo{author}{\bibfnamefont{M.}~\bibnamefont{Fabrizio}},
  \bibinfo{author}{\bibfnamefont{C.}~\bibnamefont{Castellani}},
  \bibnamefont{and} \bibinfo{author}{\bibfnamefont{E.}~\bibnamefont{Tosatti}},
  \bibinfo{journal}{Phys. Rev. Lett.} \textbf{\bibinfo{volume}{93}},
  \bibinfo{pages}{047001} (\bibinfo{year}{2004}).

\end{thebibliography}

\end{document}